\documentclass[twocolumn,prb,aps]{revtex4}

\usepackage[dvips]{graphicx}

\newcommand{\firma}[1]{\texttt{#1}}
\def\be{\begin{equation}}
\def\ee{\end{equation}}
\def\bc{\begin{center}}
\def\ec{\end{center}}
\begin{document}
\draft
\title{Ising magnets with mobile defects}
\author{W. Selke \$\ \thanks{\firma{e-mail: selke@physik.rwth-aachen.de}} , V.\ L. Pokrovsky \dag , B. B\"uchner \ddag , and T. Kroll \ddag }
\address{
\$ Institut f\"ur Theoretische Physik, Technische Hochschule, D--52056 Aachen, Germany\\
\dag Department of Physics, Texas A\&M University, College Station, Texas, 77843-4242, USA\\
\ddag II. Physikalisches Institut, Technische Hochschule, D--52056 Aachen, Germany }

\begin{abstract}
Motivated by recent experiments on cuprates with low-dimensional magnetic
interactions, a new class of two--dimensional Ising models with short--range
interactions and mobile defects is introduced and studied. The non--magnetic
defects form lines, which, as temperature increases, first meander and then
become unstable. Using Monte Carlo simulations and analytical low-- and
high--temperature considerations, the instability of the defect stripes is
monitored for various microscopic and thermodynamic quantities in detail for a
minimal model, assuming some of the couplings to be indefinitely strong. The
robustness of the findings against weakening the interactions is discussed as
well.
\end{abstract}

\pacs{05.50+q, 74.72.Dn, 75.10.Hk}
\maketitle

\section{Introduction}
Low--dimensional magnetism in high-$T_c$ superconductors
has attracted much interest, both
theoretically and experimentally \cite{Review}. In particular, striped
structures in magnets derived from the $La_2CuO_4$ compound have
been discussed rather
extensively. Motivated by related analyses and, more specifically,
by recent experiments \cite{Buch}
on $(Sr, Ca, La)_{14}Cu_{24}O_{41}$, we shall introduce
a novel class of quite simple two--dimensional Ising
models, mimicing $Cu^{2+}$ ions
by spin-1/2 Ising variables and holes by non--magnetic
defects ($S$= 0). Of course, the aim of the present study is not
to offer a full, or partial explanation of the experimental
subtleties. Indeed, beyond
the experimental motivation, the
model is hoped and believed to show various intriguing properties
being of genuine theoretical interest.

locatWe consider the situation where the spins are arranged in chains, with
antiferromagnetic interactions, $J_a$, between
adjacent chains, and a ferromagnetic
coupling, $J$, between
neighbouring spins in the chains, augmented by an
antiferromagnetic coupling, $J_0$, between next--nearest spins in the
same chain with a defect in between them. The defects are allowed
to move through the crystal, along the chains.

The defects tend to form stripes, perpendicular
to the chains, which, for increasing temperature, first
meander and then become unstable. To identify
the impact of the defect mobility on the stripe instability, a 'minimal
model' is proposed by assuming indefinitely strong interactions
in the chains, $J$ and $J_0$. This
model is studied analytically, at low and high
temperatures, and, for a wide range of temperatures, by
using standard Monte Carlo techniques \cite{Binder}. Contact
will be made to
well known descriptions of wall instabilities in two dimensions, as
have been put forward, for instance, in the context of incommensurate
superstructures in two dimensions \cite{Pokro} and step roughening
on vicinal surfaces \cite{Villain}. Deviations from these
standard scenarios will be discussed.

To study the robustness of the properties
of the minimal model and to identify other possibly interesting
aspects of this class of Ising
models as well, we also considered cases with finite couplings
in the chains. In addition, both
for the minimal model and the 'full model', the effect of an
external magnetic field has been investigated.

The paper is organized accordingly. In the next section, we
introduce the model and elucidate its experimental background. Then, we
present our results on the minimal model, followed by a
discussion on properties of the full model. A short summary
concludes the article.

\section{Models and methods}

We consider Ising models on a square lattice, setting the lattice constant
equal to one. Each lattice site $(i,j)$ is occupied either by a spin, $S_{i,j}=
\pm 1$, or by a defect corresponding to spin zero, $S_{i,j}= 0$. The defects
are assumed to be mobile along one of the axes of the lattice, which will be
called the chain direction in the following. Neighbouring sites of the same
chain, $(i,j)$ and $(i \pm 1,j)$, are not allowed to be occupied both by
defects (short range repulsion between defects). The mobility of a defect may
be influenced by a pinning potential, which, however, will be disregarded in
most of the following analysis. Even for vanishing pinning, the defect cannot
diffuse freely, in general, because an elementary move, characterized by
exchanging a defect and a, possibly flipped spin at neighbouring sites, is
affected by the magnetic interactions along and perpendicular to the chain
direction. We assume a ferromagnetic coupling, $J > 0$, between neighbouring
spins, $S_{i,j}$ and $S_{i \pm 1,j}$, along the chain augmented by an
antiferromagnetic interaction, $J_{0} < 0$, between those next--nearest spins
of the same chain, which are separated by a defect. Spins in adjacent chains,
$S_{i,j}$ and $S_{i, j \pm 1}$, are coupled antiferromagnetically, $J_a < 0$.
Accordingly, the Hamiltonian of the model may be written as

\begin{eqnarray}
{\cal H} = -\sum\limits_{ij} (J S_{i,j} S_{i \pm 1,j}
  + J_0 S_{i,j} S_{i \pm 2,j} (1- S_{i \pm 1,j}^2) \nonumber \\
  + J_a S_{i,j} S_{i,j \pm 1})
  - H \sum\limits_{ij} S_{i,j}
\end{eqnarray}
where we included a field term. Note that the defects, $S_{ij}= 0$, are
separated, along the chains, by at least one spin. We shall assume that the
number of defects is the same in each chain, determined by the defect
concentration $\Theta$, denoting the total number of defects divided by the
total number of sites.

The model describes, among others, the thermal distribution of defects, leading
to rather intriguing properties, as will be shown below. In particular, at low
temperatures, the defects tend to form stripes perpendicular to the chains
which become unstable at higher temperatures. The model is believed to be of
genuine theoretical interest.

The theoretical model may be motivated by rather recent experimental findings
for the so called telephone number compound $(Sr, Ca, La)_{14}Cu_{24}O_{41}$
which contains two magnetically one-dimensional elements. One subsystem is a
sheet like arrangement of $Cu_2O_3$ two-leg ladders, which is irrelevant in
the context of the present paper. The second subsystem is an array of $CuO_2$
chains formed by edge sharing $CuO_4$ plaquettes. For this
bond geometry with a $Cu-O-Cu$ bond angle close to 90 degree the
Goodenough-Kanamori-Anderson rules predict a ferromagnetic exchange between
nearest neighbour $Cu$ ions with spin $S=\frac{1}{2}$. This is confirmed
by neutron diffraction studies of
the magnetic structure in the ordered state which, in addition, show an
antiferromagnetic coupling of the spins perpendicular to the
chains~\cite{Matsuda98}. To our knowledge the absolute value of the
ferromagnetic coupling constant $J$ in the undoped chains has not been
determined yet from inelastic neutron data. A mean field treatment of the
magnetic susceptibility suggests a coupling constant of several
meV \cite{Carter,Buch}.

A long range magnetic order of the $Cu$ spins is only observed for certain
compositions of the telephone number compound. For many compositions the chains
contain a large number of hole-like charge carriers. These holes imply
non-magnetic $Cu$ sites \cite{Carter,Ammerdimer,KataevPRB} which inhibit the
formation of a long range ordered magnetic state. For example, in the
stoichiometric compound $Sr_{14}Cu_{24}O_{41}$ about 60 percent
of the $Cu$ sites
in the chains are non-magnetic~\cite{Osafune97,Regnault,Nucker00} and the
remaining spins form nearly independent dimers~\cite{Regnault,Ammerdimer}. The
analysis of this dimer state shows a rather large antiferromagnetic coupling
$|J_0| \simeq 11$~meV between the two $Cu$ spins
adjacent to a non-magnetic $Cu$
site~\cite{Regnault,Ammerdimer,Remark1}. This coupling is about one order of
magnitude larger than the antiferromagnetic coupling between $Cu$ ions in
adjacent chains \cite{Regnault}.

The experimental data mentioned so far yield at least three relevant magnetic
coupling constants in the telephone number compounds: a ferromagnetic coupling
$J$ of several meV between nearest neighbour $Cu$ ions in the chain, the
antiferromagnetic $|J_0| \simeq 11$~meV for $Cu$ spins adjacent to
holes~\cite{Remark1} and an antiferromagnetic
coupling $J_a$ between $Cu$ spins
in adjacent chains which is of the order of 1 meV~\cite{Regnault}. For undoped
chains the magnetic properties depend mainly on $J$ and $J_a$, whereas the
behaviour for large hole content is determined by a single coupling constant,
the antiferromagnetic exchange $J_0$. For small hole concentrations all three
magnetic interactions and their interplay should be relevant. Experimentally
such a situation is realized in $La_5Ca_9Cu_{24}O_{41}$ where the hole
content in the chains amounts to about 10 percent \cite{Nucker00}. Studies
of this compound reveal a very unusual suppression of
the magnetic order in external
fields which can not be explained in terms of conventional spin
models \cite{Buch}. It is tempting to attribute this strange behaviour to a
magnetic field induced movement of the charge carriers which frustrates the
antiferromagnetic interchain coupling.

As will be shown below our numerical results for the model Eq. (1) indeed
reveal a movement of holes in external fields. We mention that the treatment
within an Ising model is also related to experimental findings for the
telephone number compound. Different experimental data for lightly doped chains
show a strong Ising-like anisotropy~\cite{Buch,KataevPRL} which was predicted
by two independent theoretical treatments for spin chains formed by edge
sharing $CuO_4$ plaquettes \cite{Aha}.

To study the above Ising model with mobile
defects, Eq. (1), we applied analytical low
and high temperature considerations, and
Monte Carlo techniques monitoring various thermodynamic and
microscopic properties. In the simulations, we took into
account flips of single spins as well as hops of a defect
to a neighbouring site in the chain leaving a spin at the
former defect site. Of course, simulations are performed on
finite lattices with $L \times M$ sites ($L$ refers to the chain
direction). Usually, we employed full periodic boundary conditions. In
a few selected cases free boundary conditions perpendicular to
the chain direction were applied, corroborating the
results for periodic boundary conditions to
be presented in the following. To investigate
finite size effects, the linear dimensions, $L$ and $M$, were varied from 20
to 160. Typically, runs of at least $10^6$ Monte Carlo steps per spin were
performed, averaging then over a few of such realizations to estimate
error bars. The concentration of defects, $\Theta$, ranged from zero to 15
percent. In most cases, we set $\Theta =0.1$.

Physical quantities of interest include the specific heat, $C$, determined
from energy fluctuations and the temperature dependence of the
energy, the magnetization per site, $m$, and the
correlation functions
parallel, $G_1(r)= (\sum\limits_{ij} \langle S_{i,j} S_{i+r,j} \rangle)/LM $,
 and perpendicular, $G_2(r)=
 (\sum\limits_{ij} \langle S_{i,j} S_{i,j+r} \rangle)/LM $, to the
chain direction. We also calculated
other microscopic quantities describing the stability of
the defect stripes and the ordering of the spins and defects in the
chains. In particular, we computed the average
minimal distance, $d_m$, between each
defect in chain $j$, at position $(i,j)$, and those in the next
chain, at $(i',j+1)$ (i.e.
 $\sum \langle \min |i- i'| \rangle$, dividing this sum over the
defects by their number), the cluster distribution, $n_d(l)$, denoting
the probability of a cluster with $l$ spins of equal
sign in a chain (in analogy to the distribution of
cluster lengths in percolation theory \cite{Stauffer}), and the
normalized number of sign
changes of neighbouring spins, $n_c$, in the
chains. Finally, it
turned out to be quite useful to visualize the microscopic spin
and defect configurations as encountered during the simulation.

In case of the magnets mentioned above, the absolute values of both
$J_0$ and $J$ are large compared to the coupling between
chains, $J_a$. To describe then the behaviour of
the Hamiltonian, Eq. (1), at low temperatures one may consider
a simplified model in which the
spins form intact clusters in the chains between two consecutive
defects changing the sign at the defect, i.e. $J$ and
$|J_0|$ are assumed to be indefinitely strong. Quantities can now
be expressed in terms of $k_BT/|J_a|$. The
thermal excitations are due to motion of the defects
in the chains. Again, defects are
separated by at least one spin. The analysis
of this 'minimal model' will be presented in the next section.

\section{Properties of the minimal model}

\subsection{Monte Carlo results}

\begin{figure}
\begin{center}
\includegraphics[width=1.0\columnwidth]{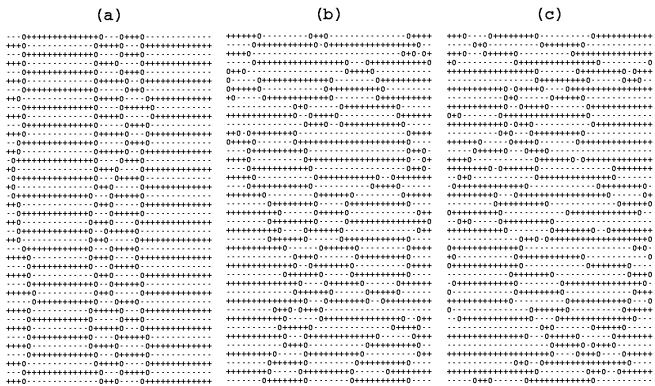}
\end{center}
\caption{Typical Monte Carlo equilibrium configurations
of the minimal model, $\Theta= 0.1$and $H= 0$, of
size $L= M=$ 40 at temperatures
$k_BT/|J_a|$= 0.6 (a), 2.6 (b), and 4.0(c).}
\end{figure}

\begin{figure}
\begin{center}
\includegraphics[width=0.8\columnwidth]{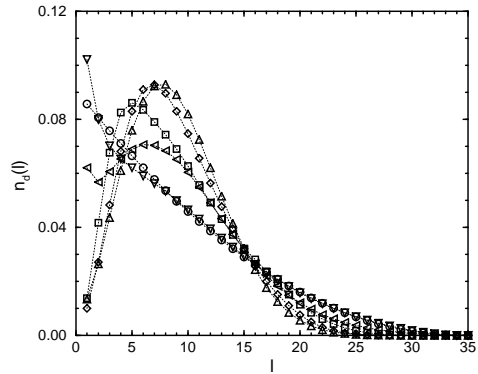}
\end{center}
\caption{Cluster distribution $n_d(l)$ at $k_BT/|J_a|$= 0 (circles), 0.5
 squares, 1.0 (diamonds), 1.5 (triangles up), 2.5 (triangles left),
and 4.0 (triangles down), for the minimal model, $\Theta= 0.1$, and
$H= 0$, of size $L= M= 40$. Results
have been obtained by exact enumerations at zero temperature, and by
simulations otherwise.}
\end{figure}

In the ground state of the minimal model with vanishing
external field, $H= 0$, the defects form straight lines
perpendicular to the chains and separating antiferromagnetic
domains of spins. The ground state
is highly degenerate, with the degeneracy depending on the
concentration $\Theta$ of defects. Each arrangement of straight defect
lines, with a separation distance between the lines
of at least two lattice spacings, has the
same, lowest possible energy, resulting in a fast decay of the
correlations $G_1$ parallel to the chains, while the spins are perfectly
correlated perpendicular to the chains. The degeneracy may be (partly)
lifted, for instance, by introducing a pinning potential or by applying
an external field as will be discussed briefly below.

Increasing the temperature, $T > 0$, the defects are
allowed to move so that the stripes start to meander
and finally break up, as exemplified in typical
Monte Carlo configurations depicted in Fig. 1. It seems
plausible that the destruction of the defect stripes by
thermal fluctuations is accompanied by singular behaviour
of thermodynamic quantities, like the specific heat, and various
correlations functions. This suggestion is, indeed, supported
by the numerical evidence discussed below. The
effect of both phenomena, meandering and breaking
up of the stripes, on various physical quantities are shown
in Figs. 2 to 6, for the case $\Theta= 0.1$. Note that in most
of the figures we did not include error bars being, typically, not larger
than the size of the symbols.

At low temperatures, deviations from the straight stripes may be characterised
by kinks and kink--antikink pairs \cite{Pokro,Villain,Pers,Selke}. Actually,
the minimal model resembles closely a terrace--step--kink (TSK) model
describing step fluctuations on vicinal surfaces. The energies of the
elementary excitations may be readily calculated. For instance, a kink with
depth of one lattice spacing costs $-J_a$, a kink--antikink pair, created by
moving a single defect by one site away from the perfect stripe, costs $-2
J_a$, for a further diffusion of that defect by another lattice spacing away
from the stripe an additional $-4 J_a$ is needed, etc. The fact that
consecutive defects in a chain cannot be closer than the minimum distance of
two lattice spacings leads to the well known phenomenon of 'entropic repulsion'
\cite{Mullins,Pers} between meandering neighbour stripes. Due to the entropic
repulsion, the meandering stripes tend to approach their average distance as
given by the concentration of defects, $\Theta$ (here, at $\Theta= 0.1$, the
average distance is ten lattice spacings). This feature is seen, e.g., in the
thermal behaviour of the cluster distribution, $n_d(l)$, where the distance
between two neighbouring defects in a chain is, in the minimal model, equal to
$(l+1)$. As shown in Fig. 2 for the moderate size $L= M= 40$, at zero
temperature, $n_d(l)$, calculated numerically by taking into account all
degenerate ground states, decreases monotonically with the separation distance
$l$. When turning on the temperature, the maximum of $n_d(l)$ moves towards
equidistant spacing of defects, here $l= 9$, and the shape of $n_d(l)$ may be
approximated by a Gaussian or Wigner function, as has been discussed recently
in the context of terrace width distributions for vicinal surfaces
\cite{Einstein}. Increasing the temperature furthermore, $T \longrightarrow
\infty$, the defects take random positions in the chains, and $n_d(l)$
acquires, of course, again the same Poissonian form as at $T=0$ for random
distribution of defect lines. The destabilization of the stripes is indicated,
for example, by a rather rapid decrease, with temperature, of the probability
to find defects at their average spacing, i. e. $n_d(l=9)$ for $\Theta= 0.1$,
see Fig. 5.

\begin{figure}
\begin{center}
\includegraphics[width=0.8\columnwidth]{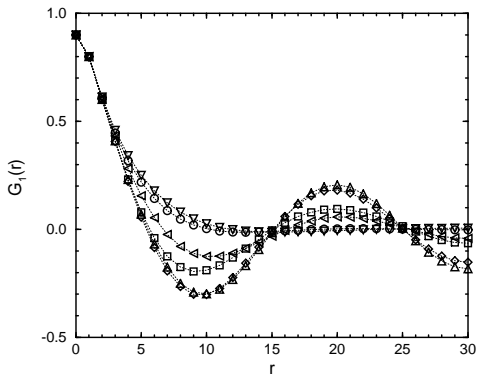}
\end{center}
\caption{Parallel correlation function $G_1(r)$ at $k_BT/|J_a|$= 0 (circles),
 0.5 (squares), 1.0 (diamonds), 1.5 (triangles up), 2.0 (triangles left),
 and 2.5 (triangles down), for the minimal model of size $L= M= 60$, as
obtained from exact enumeration, $T=0$, and simulations otherwise.}
\end{figure}

As depicted in Fig. 3, meandering and breaking up of the stripes may be also
observed in the correlation function parallel to the chains, $G_1(r)$. Again,
the behaviour at zero temperature has been determined numerically without
difficulty, for fairly small system sizes, by averaging over all ground states.
The correlations are seen to decay rapidly. The oscillations in $G_1(r)$,
already hardly visible for $L =60$, as shown in Fig. 3, become less and less
pronounced when enlarging $L$ at fixed $\Theta (=0.1)$. The asymptotics of
$G_1(r)$, for large $L$, may be determined analytically, as discussed in the
following subsection. Raising now the temperature, $T > 0$, the correlations
first become stronger, reflecting the ordering tendency which favour
equidistant stripes due to the entropic repulsion, and then decrease quite
drastically due to the thermal destabilization of the stripes. In fact, the
perpendicular correlations, $G_2$, fall off rather rapidly in the same range of
temperatures. Finally, when $T \longrightarrow \infty$, one encounters again
the behaviour at zero temperature, with the defects at random positions.

\begin{figure}
\begin{center}
\includegraphics[width=0.8\columnwidth]{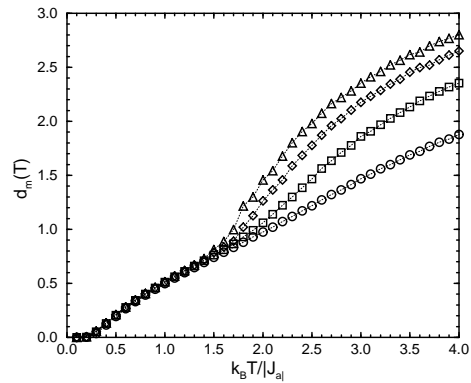}
\end{center}
\caption{Distance between defects in adjacent rows $d_m(T)$, simulating
the minimal model of size $L=M= 20$ (circles), 40 (squares), 60 (diamonds)
and 80 (triangles up).}
\end{figure}

The destruction of the defect stripes is detected directly in the average
minimal distance between defects in adjacent chains, $d_m$. Obviously, $d_m$ is
equal to zero at $T= 0$, and $d_m \approx 2 \exp(-|J_a|/k_BT)$ at low
temperatures, $k_BT \ll |J_a|$. In Fig. 4, data for various system sizes, $L=
M$ ranging from 20 to 80, are displayed. While at low temperatures, $d_m(T)$
does not depend, in fact, significantly on the system size, it starts to rise
rapidly at some characteristic temperature, with the height of the maximum in
the temperature derivative of $d_m$ increasing strongly with larger system
size. The location of the maximum, at $T_d ^{max}$, signalling the breaking up
of the stripes, moves to lower temperatures as $L$ gets larger. The
quantitative behaviour is quite similar to the one of the specific heat, to be
discussed below.

\begin{figure}
\begin{center}
\includegraphics[width=0.8\columnwidth]{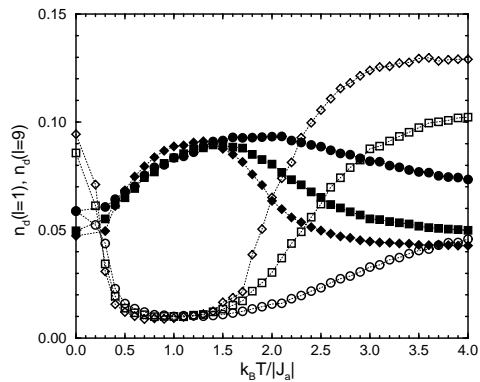}
\end{center}
\caption{Temperature and size dependence of probability for
next--nearest neighbour pairs of defects, $n_d(l=1)$ (open symbols) and
pairs at average distance $n_d(l=9)$ (full symbols), simulating systems
of size $L= M$= 20 (circles), 40 (squares), and 60 (diamonds).}
\end{figure}

One possible reason for the destabilization of the stripes are effectively
attractive interactions between neighbouring defects or lines, mediated by the
spins. Indeed, such an interaction may occur, for instance, for strongly
fluctuating stripes so that three consecutive defects in one chain, $j$, are in
the cage formed by pairs of defects in the adjacent chains, $j \pm 1$.
Consequently, two of the three defects tend to form a pair of next--nearest
neighbouring defects, as may be checked easily. In any event, the probability
of such pairs of defects is obviously given by $n_d(l=1)$. Its temperature and
size dependence is depicted in Fig. 5 (together with that of $n_d(l=9)$, as
mentioned above), showing a drastic increase close to the characteristic
temperature of the breaking up of the stripes, $T_d ^{max}$. Note that this
type of stripe instability is not included in the standard descriptions of wall
instabilities in two dimensions \cite{Pokro,Villain,Pers,Halperin}, where
either the number of walls is not fixed, giving rise to incommensurate
structures, or dislocations play a crucial role, in the context of melting of
crystals. Also the bunching of steps in TSK models with attractive step--step
interactions \cite{Weeks} or instabilities in polymer filaments due to
attractive couplings \cite{Lipo,Burk} are quite different from the loss of
stripe coherency we observe here. Of course, the breaking up of the stripes has
to be distinguished from their meandering which may result in their roughness,
driven by capillary wave excitations \cite{Villain}.

\begin{figure}
\begin{center}
\includegraphics[width=0.8\columnwidth]{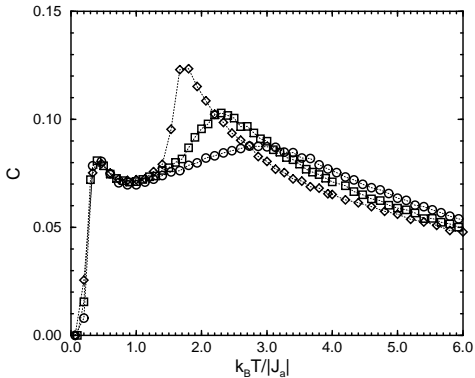}
\end{center}
\caption{Specific heat, $C$, for systems of size $L= M=$ 20 (circles),
 40 (squares), and 80 (diamonds).}
\end{figure}

\begin{figure}
\begin{center}
\includegraphics[width=0.8\columnwidth]{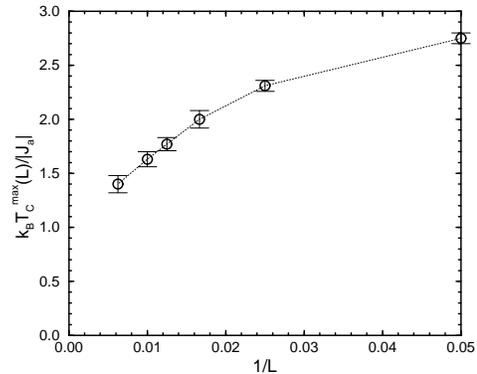}
\end{center}
\caption{Size dependence of the location of the maximum in the
specific heat, $T_C^{max}(L)$, as obtained from simulations of the
minimal model, $\Theta= 0.1$ and $H= 0$, for
 $L= M=$ 20, 40, 60, 80, 100, and 160.}
\end{figure}

Meandering and destabilization of the stripes also show up
in the specific heat, $C$, see Fig. 6 for systems
with $L= M$ sites, $L$ ranging from 20 to 80. For each size, $C$
exhibits two maxima. The
maximum at the lower temperature is almost independent of the
system size, and it is related to the kink excitations of the
stripes. The upper maximum, occurring at $T_C ^{max} (L)$, signals
the instability of the defect stripes. Its height increases
with increasing system size, indicating possibly a phase transition
in the thermodynamic limit, $L \longrightarrow \infty$. To
estimate the transition temperature, we plotted $T_C ^{max}$ versus
$1/L$, with $L$ going up to 160, see Fig. 7. From a linear extrapolation
we obtain approximately
$k_B T_C ^{max} (L = \infty)/|J_a|= 1.05 \pm 0.05$. Note that finite
size analyses, usually for sizes up to 80, for other
quantities, like $G_1$, $G_2$, and
$n_d(l=9)$, lead to similar estimates for the possible transition
temperature. However, the close agreement may be fortuituous, depending
on the type of the transition. For instance, for a
Kosterlitz--Thouless transition, the peak in the specific heat does
not occur exactly at the transition
temperature, as $L \longrightarrow \infty$. A detailed analysis
of this subtle feature is, however, beyond the scope of the
present study.

From simulations of the minimal model with $\Theta= 0.05$ and
$\Theta= 0.15$, $L=M= 40$, we infer that the characteristic
temperature, at which
the stripes become unstable, gets smaller when the concentration of
defects, $\Theta$, is increased. This observation may be explained by the
fact that the effectively attractive interactions, caused, for
instance, by the cage effect described above, may set in obviously at
lower temperature when the average distance between the stripes
decreases.

Applying an external field, $H > 0$, the ground states change
when $H$ exceeds specific critical values. For $|2 J_a| > H > |J_a|$, the
stripes are no longer straight, but they form a zig--zag
structure. In that structure, supposing the field favours
the '+' spins, each '+' cluster comprises two more spins than
the '$-$' clusters directly below and above that cluster
in the two adjacent chains. Defects bounding these '$-$' clusters are
located exactly below and above the first and last spins of
the '+' cluster. Spins and defects in each second chain
are arranged identically. Obviously, the zig--zag structures carry a
non--vanishing net magnetization. The degeneracy of the ground state
is still high, albeit somewhat smaller than in the case of
straight stripes at $H < |J_a|$, because the minimum length
of '+' clusters is now three, instead of one. For larger
fields, $H > 2 |J_a|$, the '$-$' clusters shrink
drastically: '$-$' spins occur only in the pairs of next--nearest
neighbouring defects; all other spins point in the direction of
the external field.

Monitoring various quantities at fixed
defect concentration, $\Theta= 0.1$, the stripes are observed to
become unstable at lower temperatures for
stronger fields, $0 < H < 2|J_a|$. Varying the field at fixed small
temperature, for the same defect concentration, we found, that
the destruction of the stripes seems to be accompanied by a fairly
rapid increase in the magnetization, $m$, leading to an anomaly in
the field derivative of the magnetization, similar
to experimental findings
on $(Sr,Ca,La)_{14}Cu_{24}O_{41}$ \cite{Buch}. The change
from straight to zig--zag stripes, on the other hand, leads
to a jump in $m(H)$ at $H= |J_a|$ and $T= 0$. The corresponding
maximum in the field derivative of the magnetization is, however, extremely
weak already at very low temperatures. More detailed investigations
are needed to clarify the experimental relevance of
these observations. They are beyond our present scope.

The high degeneracy of the ground states may be lifted by introducing
a pinning potential. For example, in related simulations for
$H= 0$, we found that meandering and breaking up of the
stripes seem to be qualitatively
not affected by a weak one--dimensional, along
the chain direction, harmonic regular pinning potential. More
realistically, one may introduce a random pinning potential. If it is
sufficiently strong, it is expected to destroy the defect lines
even at zero temperature. In a weak random potential, the coherency
of the defect lines gets lost or the lines collide only on a large
spatial scale, the
'Larkin length' \cite{Larkin}. At some finite temperatures, the
collision length between neighbouring lines due to thermal
fluctuations becomes smaller than the Larkin length. Starting from
this temperature, the random pinning potential can be
neglected in thermodynamics, though, it can be dominant for dynamic
phenomena. Thus, our model is thermodynamically robust with respect to
weak random potential except of a very low temperature region, in which
probably a glassy state occurs.

\subsection{Analytical results}

Here we derive the asymptotics of the spin correlation functions in the
chains, $G_1(r)= \langle S_{i,j}S_{i+r,j} \rangle$, for sufficiently
large distances $r$, first at zero
temperature, also valid at infinite temperature, and then
at non--vanishing, but small temperatures. We start with
the obvious statement that

\be
S_{i,j}S_{i+r,j}= (-1)^{n_0(i,i+r)}
\ee

\noindent
where $n_0(i,i+r)$ is the number of defects or 'zeros', in the
interval $(i,i+r)$. Eq. (2) can be rewritten as

\be
S_{i,j}S_{i+r,j}= (e^{i \pi n_0(i,i+r)} + e^{-i \pi n_0(i,i+r)})/2
\ee

\noindent
Assuming that $\langle n_0(i,i+r) \rangle = \bar{n}_0(r)$ is sufficiently
large, we apply the Gaussian statistics to the deviation
$\delta n_0(r)= n_0(i,i+r)- \bar{n}_0(r)$. Thus, the asymptotic expression
of the correlation function reads

\be
G_1(r) \approx \cos \pi \bar{n}_0(r) \exp \left(- \frac{\pi^2}{2}
\langle (\delta n_0(r))^2 \rangle\right)
\ee

The average $\bar{n}_0(r)$ is related to the concentration
of zeros $c_0= \Theta$ by
$\bar{n}_0(r)= c_0 r$. The calculation of $\langle \left(\delta n_0(r)\right)^2 \rangle$ is
more tricky. First we calculate, at zero temperature, the probability
$p(n_0,r)$ of a fixed value of $n_0$ at fixed $r$. We
find, disregarding here the value of the minimal distance
between defects,

\begin{eqnarray}
p(n_0,r)=  \exp (-r (c \ln \frac{c_0}{c} + (1-c) \ln \frac{1-c_0}{1-c}))/ \nonumber \\
 \sqrt{2 \pi r c(1-c)}
\label{prob}
\end{eqnarray}

\noindent
with $c= n_0/r$. For large numbers, the probability (\ref{prob}) has
the Gaussian form near the maximum leading to the final result:

\be
\langle (n_0 - \bar{n}_0(r)^2 \rangle= rc_0(1-c_0)
\ee

Plugging this expression into Eq. (4), we obtain the asymptotics
of $G_1(r)= \langle S_{i,j} S_{i+r,j} \rangle$ at zero temperature

\be
G_1(r, T=0) \approx \cos(\pi c_0r) \exp \left(-\frac{\pi ^2}{2}
c_0(1-c_0)r\right)
\ee

The asymptotics is valid for $r \gg r_c$, with the correlation
length $r_c= 2/(\pi ^2 c_0(1-c_0)$.
The condition of having a minimal distance of $q+1$
spacings between consecutive
defects can be taken into account by replacing $c_{(0)}$ by
$c_{(0)}/(1-c_{(0)}q)$. For our model $q=1$.

Note that the above considerations hold also
in the high--temperature limit,  $T \longrightarrow \infty$, as
mentioned before.

At finite, but small temperatures, $k_BT \ll |J_a|$, the long--distance
asymptotics of the correlation function $G_1(r)$ changes dramatically
due to the meandering of the defect stripes. This process may be
described by the free--fermion approximation \cite{Pokro}. In this
approach the lines are represented as trajectories of free
fermions. Their entropic repulsion is treated as statistical
repulsion of the fermions. The meandering of the lines means that, going
from one moment of discrete time to the next one, each fermion can
move to the left or right by one site with the amplitude (or probability)
$z= \exp (-2 |J_a|/k_BT)$. These processes are described by the Hamiltonian

\be
{\cal H} = -z\sum\limits_{i} (a^+_{i+1}a_i + a^+_ia_{i+1})
\ee

\noindent
where the fermion operators $a_i$ and $a^+_i$ obey
cyclic bondary conditions $a^{(+)}_{i+L}$= $a^{(+)}_i$.
The Hamiltonian is diagonalized by
Fourier transformation

\begin{eqnarray}
{\cal H}= -2z \sum\limits_{p} (\cos p) \alpha^+_p \alpha_p; \nonumber \\
p=2 \pi m/L (m=1,2,...L)
\end{eqnarray}

\noindent
where

\be
a_k= \sum\limits_{p} e^{ipk} \alpha_p/ \sqrt{L}
\ee

The energy band in $p$--space extends from $- \pi$ to $\pi$, but
it is filled only partly, from $-p_F$ to $p_F$, where $p_F= \pi c_0$
(to take into account the specific minimal distance
between defects, $c_0$ has to be substituted here and in the following
as before). It means that in the ground
state $\langle \alpha^+_p \alpha_p \rangle =1 $ for $|p| = p_F$, and
$\langle \alpha^+_p \alpha_p \rangle$= 0 for $p_F < |p| < \pi$. In this
approach the number of zeros between $i$ and $i+r$ is given
in terms of the fermion operators by

\be
n_0(i,i+r)= \sum_{k=i}^{i+r} a^+_k a_k
\ee

\noindent
The average
of this quantity is obviously equal to $r c_0$, as at zero
temperature, but its variance, $\langle \delta n^2 \rangle =
\langle (n_0(r) -  \bar{n}_0(r))^2 \rangle $, is drastically different. Indeed,

\be
\langle \delta n^2 \rangle= \sum_{k,k^{\prime}=i}^{i+r} \langle :a^+_ka_k: :a^+_{k^{\prime}}a_{k^{\prime}}: \rangle
\ee

\noindent
where :$XY$: denotes the normal product of the operators $X$ and $Y$.
Applying the Wick theorem, one obtains

\be
\langle \delta n^2 \rangle= \sum_{k,k^{\prime}=i}^{i+r} \langle a^+_ka_k^{\prime} \rangle \langle a_k a^+_{k^{\prime}} \rangle
\ee

\noindent
with the same summation limits. The diagonal term of this
sum, $k= k^{\prime}$, gives the contribution $r c_0(1-c_0)$, as at zero
temperature. However, it will be completely compensated by the
non--diagonal terms. Indeed, the simultaneous correlation
function for free fermions is known \cite{LL} to be equal to

\be
\langle a_k a^+_{k^{\prime}} \rangle = \sin (p_F(k-k^{\prime}))/(\pi (k-k^{\prime}))
\ee

It is easy to check that
$\langle a_k a^+_{k^{\prime}} \rangle = - \langle a^+_k a_{k^{\prime}} \rangle$. In the limit of small $p_F$, the summation in Eq.(13) can be replaced by
an integration, which can be explicitly performed without difficulty
under the additonal condition $p_Fr \gg 1$, giving

\be
\langle \delta n^2 \rangle _{nondiag}= -p_Fr/ \pi + (\ln r)/4 \pi ^2
\ee

The first term compensates the diagonal contribution. Thus, at
$0 < k_BT \ll |J_a|$, the correlation
function $G_1(r)= \langle S_{i,j} S_{i+r,j} \rangle$ is finally
approximated as

\be
G_1(r, T=0) \approx \cos(\pi c_0r)/ \sqrt{r}
\ee

Of course, the algebraic decay of the correlations is in
accordance with previous findings on free fermions in
two dimensions \cite{Pokro,Villain,Pers}. The presence
of a pinning potential is expected to establish long--range
order in the correlations. Formally, there
is no continuous change from the exponential decay
of the correlations, at $T= 0$, to the algebraic decay at non--vanishing
temperatures. However, the final expression, Eq. (16), is valid only
for a system whose size, $L$, perpendicular to the stripes exceeds the
collision length of the fermions, $l_{coll}= 1/(z c_0^2)$, which
goes to infinity as $T \longrightarrow 0$. At a fixed value of $L$, there
exists a crossover temperature $T_{cr} \approx |J_a|/\ln(L c_0^2)$ at which
the exponential decay of the correlations goes over into an algebraic
one. Note that the size, $L \gg 1/(z c_0^2)$, is rather large at
the concentration we mostly considered in the
simulations, $c_0= \Theta= 0.1$, and the crossover effect
plays no role there. However, such
sizes are not big in experimental systems.

At large temperatures, the correlations $G_1$ are believed
to decay exponentially, see
Eq. (7) which also holds at infinite temperature. Thence one
expects a transition from algebraic decay at low temperatures
to an exponential decay at high temperatures, in accordance with
the simulational results.

\section{Beyond the minimal model}

In the following, we shall present results on the full model, Eq. (1), with
finite ferromagnetic couplings, $J$, between neighbouring spins in a chain and
antiferromagnetic couplings, $J_0$, between next--nearest spins in a chain
separated by a defect. In the simulations we choose $J_0/J= -6.25$ and $J_a/J$
ranging from zero to minus one. This choice is, again, motivated by the
experimental findings mentioned above.

\begin{figure}
\begin{center}
\includegraphics[width=0.8\columnwidth]{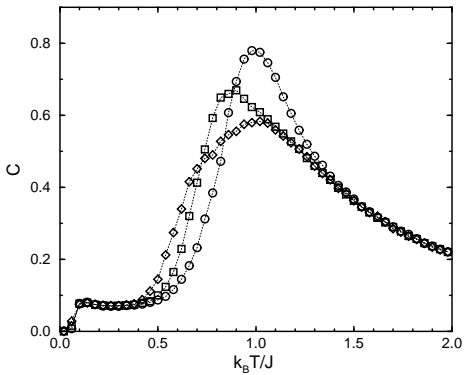}
\end{center}
\caption{Specific heat, $C$, of the full model, Eq. (1), at
$J_a/J= -0.3$, $H= 0$, and $\Theta$= 0.1, as obtained from simulations
for systems of size $L= M=$ 20 (circles), 40 (squares), and 80 (diamonds).}
\end{figure}

At zero temperature and small fields, one obtains the same highly degenerate
ground states of perfectly straight or zig--zag stripes as in the minimal
model. Likewise, the behaviour at low temperatures and $H < 2 |J_a|$ is
characterised by the meandering of the stripes as in the minimal model,
followed by the stripe instability at higher temperatures. However, the
instability may be masked, for instance, in the specific heat at $J_a/J= -0.3$,
$\Theta= 0.1$, and $H=0$, for systems of sizes up to $L$= 80, as depicted in
Fig. 8 and to be discussed in the following.

In this case, in addition to the weak, almost size--independent maximum at
low temperatures due to stripe meandering, the specific heat
displays a rather pronounced peak at higher temperatures being
much stronger than the one in the minimal model. The peak, however, gets
smaller when the system size increases. Indeed, it is
non--critical, stemming from energy fluctuations by breaking
bonds, $J$, between spins in the chains. It persists when
setting $J_a= 0$, i. e. in the one--dimensional limit exhibiting, of
course, no phase transition at all. In that limit, the lower
maximum in $C$ disappears, because there are no stripes. Actually, the
defects and their mobility play an important role in breaking the
bonds betwen spins in the chains. For instance, when
a defect moves next to a flipped spin, the spin on the other side of
the defect will be flipped rather easily, costing an energy
of 2$J$ in the ferromagnetic
coupling. In contrast, in chain without defects, an
energy of 4$J$ is needed to create the elementary
excitation comprising a pair of neighbouring spins with opposite
signs.

We also analysed, for the case depicted in Fig. 8, the simulational
data for the spin correlations, $G_1$ and $G_2$, the cluster
distribution, $n_d(l)$, and the average distance between
defects in adjacent chains, $d_m$. The data provide
strong evidence that the defect stripes become unstable at about
the same temperature, measured in
$k_BT/|J_a|$, as in the minimal model. Thence the impact
of the spin flips on the location of the stripe instability seems
to be rather small for this choice
of parameters. In principle, the spin flips
may lead to new mechanisms, different
from the effectively attractive interactions
between the defects discussed for the minimal model, to
destruct the coherency of the stripes. To detect the stripe
instability in the specific heat at $J_a/J= -0.3$, presumably
significantly larger systems have to be simulated. As shown in
Fig. 8, e.g., for $L=M$= 80 merely a shoulder in $C$ starts to
develop, at about the temperature where the minimal model shows
a peak in $C$, compare with Fig. 6. Note
that very long runs, especially for large systems, are needed to get
sufficiently good statistics for the Monte Carlo data.

Applying a magnetic field, $H > 0$, the breaking up of the
stripes is found to shift to lower temperatures, as in the
minimal model.

\begin{figure}
\begin{center}
\includegraphics[width=0.8\columnwidth]{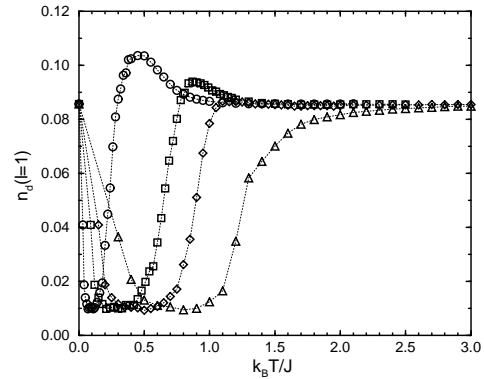}
\end{center}
\caption{Cluster distribution, $n_d(l=1)$, in the full model, Eq. (1), at
$H$= 0, $\Theta$= 0.1,and $J_a/J= -0.1$ (circles), $-0.3$ (squares),
 $-0.5$ (diamonds), and $-1.0$ (triangles), as obtained from simulations
for systems of size $L= M$= 40, see Fig. 2.}
\end{figure}

When choosing a smaller, but non--vanishing ratio
of $-J_a/J$, one gets closer to
the minimal model. We did simulations for $J_a/J= -0.1$. In fact, there
the instability of the defect stripes is also indicated by
a maximum in the specific heat, as in the
minimal model, already for rather small systems, e.g., $L= 40$, followed
by the large non--critical peak due to the spin flips in the chains.

On the other
hand, when weakening $J$ with respect to $J_a$, the stripe instability, as
indicated by the rapid increase of the minimal distance $d_m$, may occur
quite close to the pronounced, non--critical
maximum in $C$ being due to the spin flips in
the chains. In particular, for
$J_a/J= -0.5$ and, especially, $-1.0$, one then observes another clearly
visible maximum nearby the non--critical peak in $C$
already for small and moderate system sizes, e.g. for $L$= 40, due
to the stripe instabilty. The location of the instability, measured
in units of $k_BT/J$, increases with increasing
ratio $-J_a/J$. Interestingly enough, near the instability, the probability
of finding pairs of next--nearest neighbouring defects, $n_d(l=1)$, now
does not show any longer the overshooting phenomenon, compared to complete
disorder at $T \longrightarrow \infty$, in contrast to the situation in
the minimal model and for small ratios $J_a/J$, see Fig. 9. The
stripes become unstable at temperatures of the order
of the ferromagnetic spin coupling $J$, and then
the tendency to form pairs of next--nearest neighbouring defects
is diminished by thermal disordering.

\section{Summary}

In this paper a two--dimensional Ising model with defects being mobile along
the chain direction has been introduced. Albeit the model has been motivated by
recent experiments on cuprates with low dimensional magnetic interactions, the
model is believed to be of genuine theoretical interest as well.

In particular, based on analytical, asymptotical considerations
at low and high temperatures as well as on  Monte Carlo
techniques, the model is found to desribe formation of defect
stripes, their thermal meandering and, at higher
temperatures, their destabilization.

The meandering and the instability of the stripes is discussed in
the framework of a minimal model, assuming infinitely
strong couplings between the spins along the chains. The
instability is signalled by pronounced anomalies in
spin correlation functions, in the spin cluster distribution along
the chain, in the specific heat, and in the minimum distance between
defects in neighbouring chains. The breaking up of the stripes
seems to be caused by an effectively attractive interaction between
the defects mediated by the spins.

The main features of the stripe instability persist when
replacing the infinite couplings by, presumbaly, experimentally more
realistic values. However, the anomaly in the specific
may be masked for rather small systems, and thermal disordering
and spin flips may also
reduce the pairing tendency of the defects.

New experimental data on a stripe instability in
$(Sr,Ca,La)_{14}Cu_{24}O_{41}$, together with a more detailed discussion on
possible theoretical interpretations, will be presented elsewhere.

\acknowledgments
We like to thank T. W. Burkhardt, T. L. Einstein, and
D. Stauffer for useful discussions. One of us, V.L.P., acknowledges
the financial support from the Humboldt foundation and from
NSF under grant DMR 00721115 as well as the kind hospitality at
the Institute of Theoretical Physics of Aachen University.



\begin{references}
\bibitem{Remark1} The experimental data do not exhibit signatures of a large
antiferromagnetic coupling between next nearest Cu ions for chains without
holes suggesting that the holes are crucial for the size of $J_0$ as will be
shown and discussed in a forthcoming publication.
\end{references}
\end{document}